\RequirePackage{amsmath} 
\documentclass[aps,pra,onecolumn,nofootinbib]{revtex4-2}
\usepackage{graphicx}
\usepackage{dcolumn}
\usepackage{microtype}
\usepackage{cases,amssymb,amsmath}
\usepackage{xcolor}
\usepackage{bm}
\usepackage[colorlinks=true, citecolor=blue, urlcolor=blue, linkcolor=blue]{hyperref}

\newcommand{\unibath}{Department of Physics, University of Bath, Bath BA2 7AY, United Kingdom}
\newcommand{\cop}{Centre for Photonics, University of Bath, Bath BA2 7AY, United Kingdom}
\newcommand{\VTT}{VTT Technical Research Centre of Finland, FI-02044 Espoo, Finland}
\newcommand{\NPL}{National Physical Laboratory, Hampton Road, Teddington TW11 0LW, England, United Kingdom}

\begin{document}

\title{Multiple soliton regimes enabled by parametric down-conversion in $\protect\chi^{(2)}+\protect\chi^{(3)}$ microresonators}

\author{Francesco Rinaldo Talenti}
\affiliation{\unibath}
\affiliation{\cop}

\author{Tommi Isoniemi}
\affiliation{\unibath}
\affiliation{\VTT}

\author{Jonathan Silver}
\affiliation{\NPL}

\author{Dmitry Skryabin}
\affiliation{\unibath}
\affiliation{\cop}
\affiliation{\NPL}

\begin{abstract}

Half-harmonic generation in microresonators with mixed $\chi^{(2)}$ and $\chi^{(3)}$ nonlinearities provides access to a rich landscape of two-colour soliton states and frequency combs. We numerically demonstrate the formation of both bright parametric and topological two-colour solitons at relatively modest pump powers and phase-mismatch parameters. At higher powers and larger phase mismatches, we identify a regime where parametric down-conversion assists dark-soliton formation in the normally dispersive pump field. While the dark-soliton core is primarily shaped by the Kerr nonlinearity, nonlinear coupling between the pump and its half-harmonic modes enables soliton excitation using a standard laser-frequency scanning technique, a capability not normally associated with dark solitons. The simulations use parameters representative of lithium-niobate microresonators pumped near 775 nm and generating a half-harmonic near 1550 nm. These results reveal how the interplay of $\chi^{(2)}$ and $\chi^{(3)}$ nonlinearities can substantially expand the range and types of experimentally accessible soliton states in a single integrated frequency-comb platform.
\end{abstract}


\maketitle

\section{\label{sec:1} Introduction}

Coherent optical frequency combs (OFCs) generated in integrated photonic circuits and associated with dissipative solitons have profoundly impacted modern metrology, spectroscopy, and precision timing \cite{ Bellini:1998, Diddams:2000, didrev}. Among the various microresonator platforms for OFC generation, optical parametric oscillators (OPOs) based on quadratic, $\chi^{(2)}$, nonlinearity \cite{ulvila2013, Ricciardi2015, Mosca2018, iolandarev,jank,boes} occupy a distinct niche relative to the extensively studied Kerr-only, $\chi^{(3)}$, microresonators 
\cite{Savchenkov:2004,Kippenberg:2004,Haye:2007Nat, kiprev, Herr:2014NATPHOT,Haye:2011PRL}. In recent years, dissipative solitons supported by $\chi^{(2)}$ nonlinear interactions have emerged as a rapidly developing research area, revealing a diversity of soliton states that extends well beyond the conventional Kerr-soliton paradigm \cite{nie,skrrev}.

The origin of this richness lies in the intrinsically multicomponent nature of $\chi^{(2)}$ solitons. The pump field and its second harmonic are separated by an octave in frequency, which naturally leads to substantial differences in group velocities and makes opposite signs of group-velocity dispersion at the two soliton components a common situation rather than an exception \cite{tal25}, pushing researchers to conceive innovative architectures capable of tailoring dispersion across broader frequency spectra \cite{moille2018OL, moille2023fourier,Tal22, lucas2023tailoring, tal26}. 

Achieving low-threshold, frequency-tunable, octave-spanning frequency combs remains a major driving force behind the development of $\chi^{(2)}$ microresonator platforms \cite{boes}. At the same time, higher-power operating regimes are of considerable interest in their own right. In these regimes, Kerr nonlinear effects become unavoidable \cite{ he2019, vil19,bruch2021,ding2024,simult, Bengel2026OL, Sun26}, and their interplay with quadratic nonlinear interactions can be controlled through the pump power and phase-matching conditions \cite{tal25}. Understanding this mixed $\chi^{(2)}+\chi^{(3)}$ regime is therefore essential for describing the full range of nonlinear states supported by microresonator OPOs.

Modelling systems with mixed nonlinearities presents its own challenges. One widely used approach eliminates one of the interacting fields by expressing it as a convolution integral and substituting the result into the remaining field equation, see, e.g.,~\cite{pedro2019,topo}.
This technique works particularly well for purely quadratic singly resonant nonlinear resonators. Eventually, when a non-negligible part of the driving field is converted to one or several different spectral carriers, a multi-envelope mean field approach can be adopted \cite{phil24}. This framework efficiently describes nonlinear mean-field coupling in both up- \cite{tal25OL, Bengel2026OL} and down-conversion \cite{Sun26, he2019} configurations. More generally, it can model any kind of nonlinear wave coupling, such as recently demonstrated in cross-polarised Kerr-coupled systems \cite{hill2020effects, lucas2025faticons}.

An alternative approach relies on coupled-mode equations, which naturally accommodate the simultaneous presence of $\chi^{(2)}$ and $\chi^{(3)}$ nonlinearities~\cite{bruch2021,josab,puz2}. In addition to its modelling flexibility, the coupled-mode formalism provides a direct connection to experimentally measured discrete comb spectra and individual mode amplitudes.
In highly multimode systems like OFCs, different sidebands compete simultaneously across various orders of both $\chi^{(2)}$ and $\chi^{(3)}$. A coupled-mode theory (CMT) naturally embeds this information, offering direct access to the underlying dynamics and intricate threshold mechanisms.
CMT has already revealed a ladder-like sequence of instabilities leading to OPO signal generation in the few-mode regime, with theoretical predictions subsequently confirmed experimentally in microresonators~\cite{puz22,ingo1,ingo2}.

 In this Letter, we employ a CMT to investigate a variety of soliton-generation regimes in microresonator OPOs and map their existence domains in the global pump-power versus phase-mismatch ($\varepsilon_0$) parameter space.

In this work, we investigate soliton excitation within a thin-film lithium 
niobate (LN) microresonator, as it is widely recognized as one  of the most 
efficient $\chi^{(2)}$ platforms \cite{luo2019modalphsematching, lu2019, wolf2018quasi, yang2024}. The system is driven by a pump at $\lambda_p=775\ \mathrm{nm}$ 
operating in the normal group-velocity dispersion regime~\cite{lonc}, while 
its half-harmonic component at $\lambda_{hh}=1550\ \mathrm{nm}$ experiences 
anomalous dispersion. We study this excitation using a standard 
laser-frequency scanning technique.
\begin{figure*}[t]
\includegraphics[width=\textwidth]{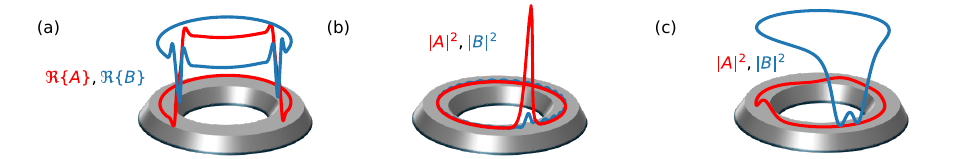}
  \caption{Different localised states found and discussed in this letter: (a) topological, (b) two-colour bright and (c) dark soliton states.}
  \label{Fig1:scheme}
\end{figure*}

We identify $\varepsilon_0$ as the pivotal parameter determining the relative positioning of the parametric threshold boundaries. It thus distinguishes different nonlinear regimes, where either $\chi^{(2)}$ and $\chi^{(3)}$ effects compete  or one of the two prevails.
Because parametric down-conversion power thresholds are orders of magnitude lower than their Kerr counterparts, configurations close to the matching condition $\varepsilon_0 \sim 0$ are predominantly governed by second-order effects. Under these conditions, the HH homogeneous state naturally evolves into a solution with a $\pi$-phase jump \cite{oppo99,oppo01,pedro2019,old1}. This discrete phase shift of the HH field triggers localized states at the pump wavelength (Fig. \ref{Fig1:scheme} (a)) resulting in a  topological two-colour soliton.
Our results echo recent topological OPO-soliton experiments performed in a markedly different lithium-niobate photonic-integrated-circuit platform, where a low-repetition-rateresonator (10MHz versus our 0.5THz) was used to provide feedback for the HH field, while the 
pump field was filtered out of the resonator loop \cite{topo}.


For sufficiently large $|\varepsilon_0|$, the $\chi^{(2)}$ and $\chi^{(3)}$ effects start competing. In particular, for $\varepsilon_0\ll 0$ we uncover  families of two-colour bright solitons (see Fig. \ref{Fig1:scheme} (b)). The existence of these states is notable in view of the opposite dispersion signs experienced by the pump and HH fields, suggesting that the experimental observations of HH bright solitons reported in Refs.~\cite{bruch2021,tal25OL}, where both frequencies operated in the anomalous-dispersion regime, represent only a subset of the accessible operating conditions.

For $\varepsilon_0 \gg 0$ and at higher pump powers, we uncover a regime of dark-soliton generation~\cite{dark1, dark2, dark3} (Fig. \ref{Fig1:scheme} (c)). This dynamics is predominantly governed by Kerr, $\chi^{(3)}$, effects. Importantly, the weak $\chi^{(2)}$-induced coupling to the HH field plays a crucial role in the excitation process of the pump-only soliton. In contrast to Kerr-only microresonators, where dark-soliton generation typically requires cross-mode interactions or external triggers, the present system supports robust excitation via conventional frequency scans.

\section{The model}

We consider an LN periodically-poled ring resonator \cite{Zhu2021LN}, assuming that the pump field at frequency $\omega_p$ couples to a resonator mode with number $N_b$ and frequency $\omega_{0b}$. The HH frequency, $\omega_p/2$, lands near the mode $N_a$ with frequency $\omega_{0a}$. The resonator can be periodically poled with a radial pattern of electrodes for z-cut materials, and if $Q$ is the number of periods along the ring circumference, then momentum conservation requires 
\begin{equation}
2N_a+ Q=N_b. \label{mom}
\end{equation}

Representing the pump, $B$, and the half-harmonic, $A$, envelopes as sums over the microresonator modes,
\begin{equation}
A=\sum_\mu a_\mu(t)e^{i\mu\vartheta},\ \ \ B=\sum_\mu b_\mu(t)e^{i\mu\vartheta},~\mu=0,\pm 1,\pm 2,\dots,
\label{Eqmodes}
\end{equation}
the corresponding electric fields are reconstructed as
$Ae^{iN_a\vartheta-i\tfrac{1}{2}\omega_pt}+c.c.$ and
$Be^{iN_b\vartheta-i\omega_pt}+c.c.$.
Here $\vartheta=(0,2\pi]$ is the angular coordinate in the laboratory reference frame.

The resonance frequencies of the two groups of modes participating in the frequency conversion are defined as $\omega_{\mu a}=\omega_{0 a}+D_{1a}\mu+\tfrac{1}{2}D_{2a}\mu^2$ and $\omega_{\mu b}=\omega_{0 b}+D_{1b}\mu+\tfrac{1}{2}D_{2b}\mu^2$, where $D_{1a,1b}$ and $D_{2a,2b}$ are the corresponding repetition rates and dispersion parameters, respectively. Higher-order dispersion terms are neglected. 
The resonator is assumed to have a radius of 70$\mu$m, a ridge width of 1.8$\mu$m, and an etching depth of $410$ nm in the $590$ nm deep lithium niobate layer~\cite{lu2023}, yielding the dispersion values listed in Table 1.

\begin{table}[t]
\centering
\begin{tabular}{cccccc}
\hline\hline
$\gamma_{2a}/2\pi$ & $\gamma_{2b}/2\pi$ & $\gamma_{3a}/2\pi$ & $\gamma_{3b}/2\pi$ & $v$ & 
\\ \hline
4 GHz/$\sqrt{W}$& 8 GHz/$\sqrt{W}$& 
2.5 MHz/W& 5 MHz/W &368&
\\ \hline\hline
 $D_{1a}/2\pi$ & $D_{1b}/2\pi$ & $D_{2a}/2\pi$ & $D_{2b}/2\pi$  
 & $\kappa_a/2\pi$ & $\kappa_b/2\pi$ \\
\hline
286 GHz& 289 GHz& 14 MHz & -18 MHz& 300 MHz& 600 MHz\\
\hline\hline
\end{tabular}
\caption{\label{tab1} Resonator  parameters: $\gamma_{2a,2b}$ and $\gamma_{2a,2b}$ are nonlinear coefficients characterising quadratic and Kerr nonlinearities, respectively; $v$ is the resonator power buildup factor; $D_{1a,1b}$ are the repetition rates; $D_{2a,2b}$ are dispersion coefficients; $\kappa_{a,b}$ are linewidth parameters.   }
\end{table}

Our focus here is on multimode generation and soliton mode-locking mechanisms stemming from the down-conversion of pump photons with frequency $\omega_p$ into pairs of signal photons, $\omega_p\rightarrow \omega_{\mu_1 a}+\omega_{\mu_2 a}$, where $\mu_p=\mu_1+\mu_2$. The corresponding phase-matching parameter, defined by the resonator design and temperature, is
\begin{equation}
\varepsilon_{\mu_1,\mu_2}=\omega_{\mu_1 a}+\omega_{\mu_2 a}-\omega_{\mu_p b}\ \ \ .
\label{Eqeps}
\end{equation}

Nonlinear interaction between the modes is mediated by the interplay of quadratic, $\chi^{(2)}$, and Kerr, $\chi^{(3)}$, nonlinearities and is governed by the coupled-mode equations \cite{josab}:
\begin{align}
    i\partial_t a_\mu=& (\delta_{\mu a}- \tfrac{1}{2} i\kappa_a)a_\mu  -\gamma_{2a} \sum_{\mu_1}b_{\mu_1}a_{\mu_1-\mu}^*\nonumber\\ -& \gamma_{3a}\sum_{\mu_1,\mu_2}
    a_{\mu_1}a_{\mu_2}a_{\mu_1+\mu_2-\mu}^*-2  \gamma_{3a}    \sum_{\mu_1,\mu_2}a_{\mu_1}b_{\mu_2}b_{\mu_1+\mu_2-\mu}^*
    \label{Eqa}\\
    i\partial_t b_\mu=& \delta_{\mu b} b_\mu -i\tfrac{1}{2}\kappa_b\left(b_\mu -\widehat\delta_{\mu,\mu_p}{\cal H} \right)-\gamma_{2b}\sum_{\mu_1}a_{\mu_1}a_{\mu-\mu_1}\nonumber\\ -
    & \gamma_{3b}\sum_{\mu_1,\mu_2}
    b_{\mu_1}b_{\mu_2}b_{\mu_1+\mu_2-\mu}^*-2  \gamma_{3b}    \sum_{\mu_1,\mu_2}b_{\mu_1}a_{\mu_2}a_{\mu_1+\mu_2-\mu}^*
    \label{Eqb}
\end{align}
Here $\gamma_{2a,2b,3a,3b}$ are the nonlinear parameters, $\kappa_{a,b}$ are the linewidths, and $\mathcal{H}=\sqrt{v\mathcal {W}}$ is the pump parameter, where $v$ is the resonator power buildup factor, see Table \ref{tab1}, and which assumes critical coupling~\cite{lu2023}. $\cal W$ is the laser power in the bus waveguide, and $\widehat h\delta_{\mu,\mu_p}$ is the Kronecker delta function.
Switching to a reference frame rotating at the rate $D_{1a}$, the detuning parameters are defined as
\begin{align}
\delta_{\mu a}=\omega_{\mu a}-\tfrac{1}{2}\omega_p-D_{1a}\mu, ~\delta_{\mu b}=\omega_{\mu b}-\omega_p-D_{1a}\mu. \label{Eqdelta}
\end{align}

\noindent
Two distinct phase-matching scenarios arise depending on whether an even- or odd-numbered mode is pumped~\cite{ingo1}. In this work, we restrict our attention to the even-mode pumping case, corresponding to $\mu_p=0$, $\mu_{1}=\mu$, and $\mu_{2}=-\mu$, for which
\begin{eqnarray}
&&\varepsilon_{\mu_1,\mu_2}=\varepsilon_{\mu,-\mu}=\varepsilon_0+\mu^2 D_{2a}, 
\label{ep}\end{eqnarray}
where, using Eq.\ref{mom}, the residual phase mismatch at $\varepsilon_0$ can be written as
\begin{eqnarray}
&&\varepsilon_0=2\omega_{0a}-\omega_{0b}=
\frac{c}{R}
\left(\frac{2N_a}{n_{\mathrm{eff}}(\omega_{0a})}-
\frac{N_b}{n_{\mathrm{eff}}(\omega_{0b})}\right)\nonumber\\
&&=\mathrm{FSR}_b
\left[
\frac{N_b-Q}
{n_{\mathrm{eff}}(\omega_{0a})/n_{\mathrm{eff}}(\omega_{0b})}
-N_b
\right].\label{eps1}
\end{eqnarray}
Here, $c$ is the speed of light in vacuum, $R$ is the resonator radius, and $\mathrm{FSR}_b$ is free spectral range at the pump frequency. The effective refractive index for the TM polarization in our geometry is $n_{\mathrm{eff}}(\omega_{0b})\simeq 2.17$ for the $b$-field and $n_{\mathrm{eff}}(\omega_{0a})\simeq 1.87$ for the $a$-field. For a representative value of $N_a=1000$ and $Q=0$, the quantity in square brackets evaluates to approximately $140$. Thus, in an unpoled resonator, the half-momentum mode, $N_a=N_b/2$, is detuned from the half-frequency mode, $\omega_p/2$, by about $140$ FSRs and achieving phase matching, $\varepsilon_0\approx 0$, requires approximately $160$ poling periods, 
$Q\approx N_b(1-n_{\mathrm{eff}}(\omega_{0a})/n_{\mathrm{eff}}(\omega_{0b}))$.

Before proceeding further, we note that either of the two detunings appearing in Eqs.~\ref{Eqdelta} can be expressed through the phase-mismatch parameter $\varepsilon_0$ and the remaining detuning. For example,
\begin{equation}
\delta_{0b}=2\delta_{0a}-\varepsilon_0\equiv \delta,
\label{db}
\end{equation}
or
\begin{equation}
\delta_{0a}=\frac{1}{2}(\delta_{0b}+\varepsilon_0),
\label{da}
\end{equation}
so that one of the two detunings, depending on convenience, and $\varepsilon_0$ can be used as a pair of independent parameters.


\section{Cascaded generation of OPO sideband pairs}

\begin{figure*}[!b]
\includegraphics[width=\textwidth]{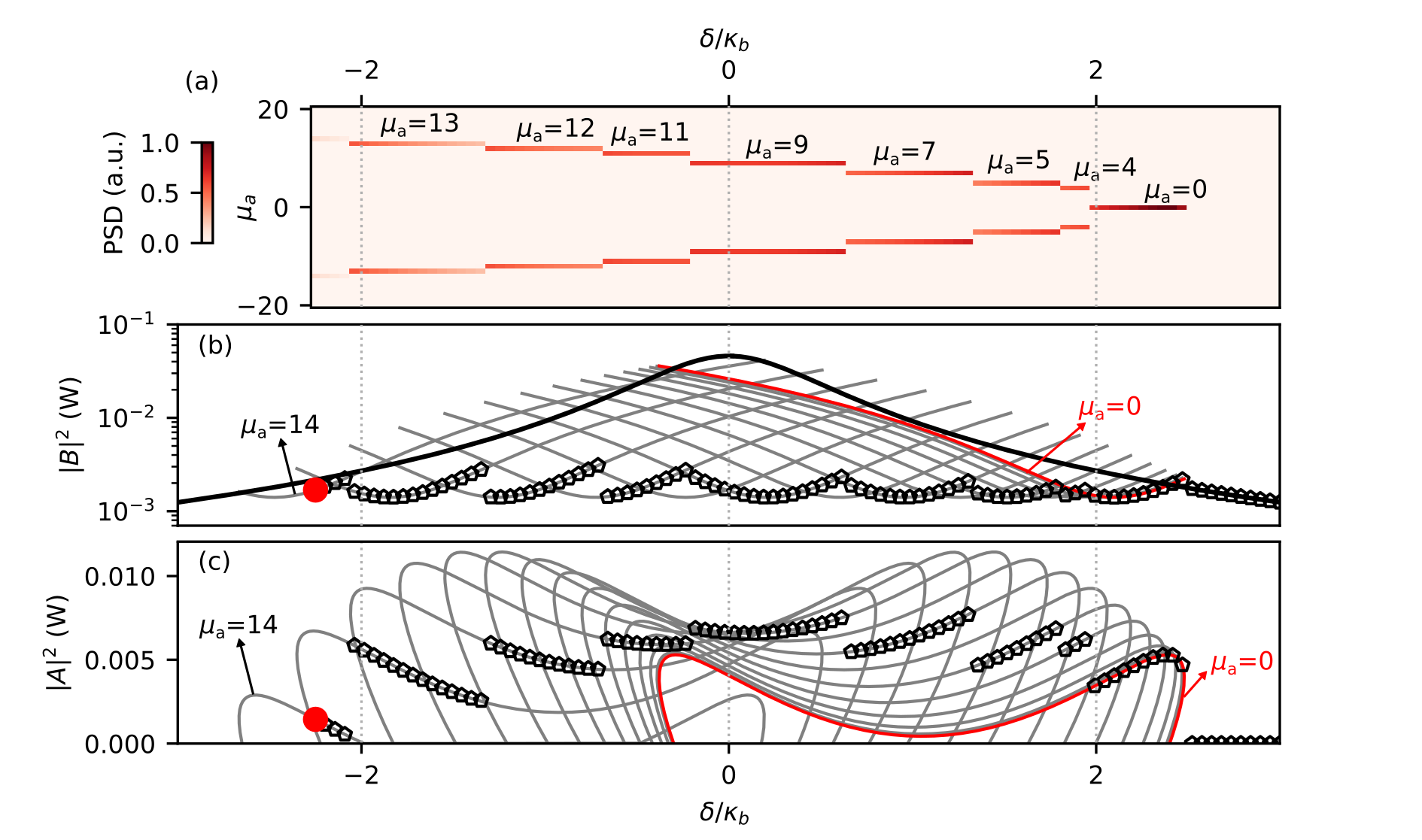}

  \caption{An example of the OPO cascade for low input power $\mathcal{W}=0.15$mW. (a) Power Spectral Density (PSD) evolution while varying the detuning. (b) Pump (log scale) and (c) half-harmonic intracavity powers. The gray lines and diamonds show the 3-mode approximation and full model numerical solutions, respectively. The red dot shows starting point of the detuning scan of Eqs. \ref{Eqa}, \ref{Eqb}. The red line highlights the degenerate OPO case, $\mu=0$. The black background line in plot (b) is the cold pump resonance, i.e. the linear homogeneous steady state.
    We consider parameter values listed in Table \ref{tab1} with a phase mismatch $\varepsilon_0=2.1 \kappa_b$.
      }
  \label{Fig2:stair_OPO}
\end{figure*}

\begin{figure*}[!t]
\includegraphics[width=\textwidth]{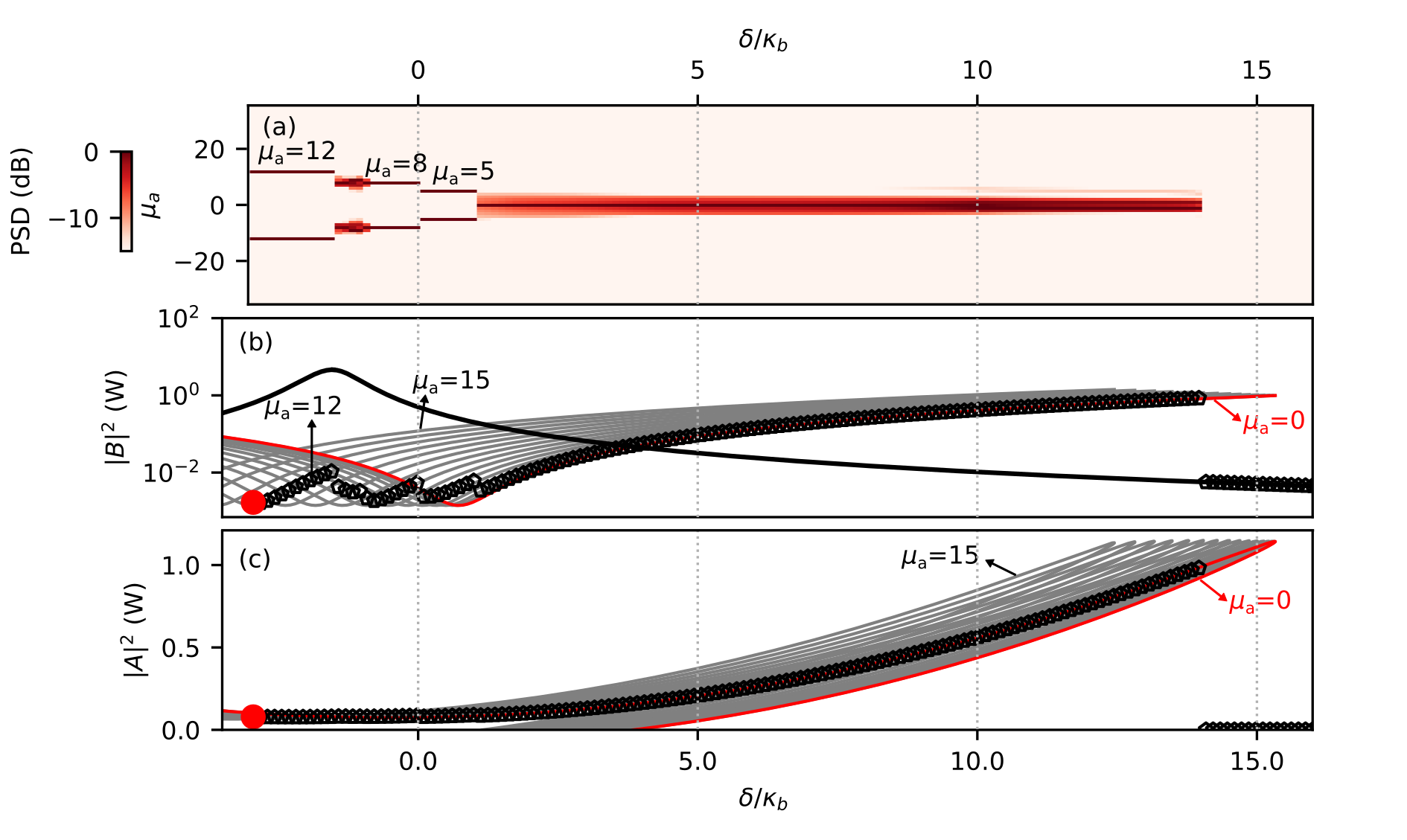}
  \caption{Ladder structures for $\mathcal{W}=15$ mW, $\varepsilon_0=2.1 \kappa_b$.
  }
  \label{Fig3:stair2}
\end{figure*}
At first sight, the classical dynamics of a microresonator OPO should be largely determined by the threshold conditions for all possible signal-idler pairs emerging from noise, i.e., by the primary Benjamin--Feir instability of the homogeneous pump state. However, recent experimental and numerical studies of purely quadratic systems reveal a considerably richer picture~\cite{puz22,ingo2}. Rather than being generated directly from the pump, successive signal-idler pairs emerge through secondary, i.e., Eckhaus, instabilities of previously established side-band states~\cite{puz22,ingo2}. The resulting hierarchy of states forms an instability cascade that structures the evolution of the OPO spectrum that follows the pump frequency scan across the resonance.

An example of such a cascade is shown in Fig.~\ref{Fig2:stair_OPO}(a), while in Fig.~\ref{Fig2:stair_OPO}(b) we report the pump component of the OPO dynamics, and Fig.~\ref{Fig2:stair_OPO}(c) presents the corresponding powers down converted to the HH side-band pair. Remarkably, despite the complexity introduced by the simultaneous presence of $\chi^{(2)}$ and $\chi^{(3)}$ nonlinearities, the entire cascade can be described analytically in the three-modes pure OPO-like limit. The derivation is provided in the Supplementary Material and shows excellent agreement with the numerical results for low enough input powers (in Fig. \ref{Fig2:stair_OPO} we consider $\mathcal{W}=150 \ \mu$W). 
As the detuning increases, the progressive generation of side-band pairs drives the system towards the degenerate OPO (dOPO) state. The direction of the cascade, i.e., if dOPO state is located left or right, is determined by the sign of the HH dispersion coefficient. For the normal-dispersion case considered here, $D_{2a}>0$, the dOPO state is shifted to the right-hand side of the cascade. Reversing the dispersion sign, $D_{2a}<0$, mirrors the structure, placing the dOPO state on its left-hand side. 

For larger pump power ($\mathcal{W}=15$ mW, Fig.~\ref{Fig3:stair2}), the corresponding OPO states become increasingly difficult to sustain. Large-amplitude power fluctuations develop within the broadened sidebands and can drive the system, at their minima, towards the non-OPO state, $a_\mu=0$. As a result, the side-band states gradually lose stability over an increasingly broad detuning range.
As the input power increases, the analytic formulation derived in the pure three-mode OPO limit loses reliability as the system begins to exhibit multimode dynamics. 



\section{Dark solitons}
\begin{figure*}[!b]
\includegraphics[width=\textwidth]{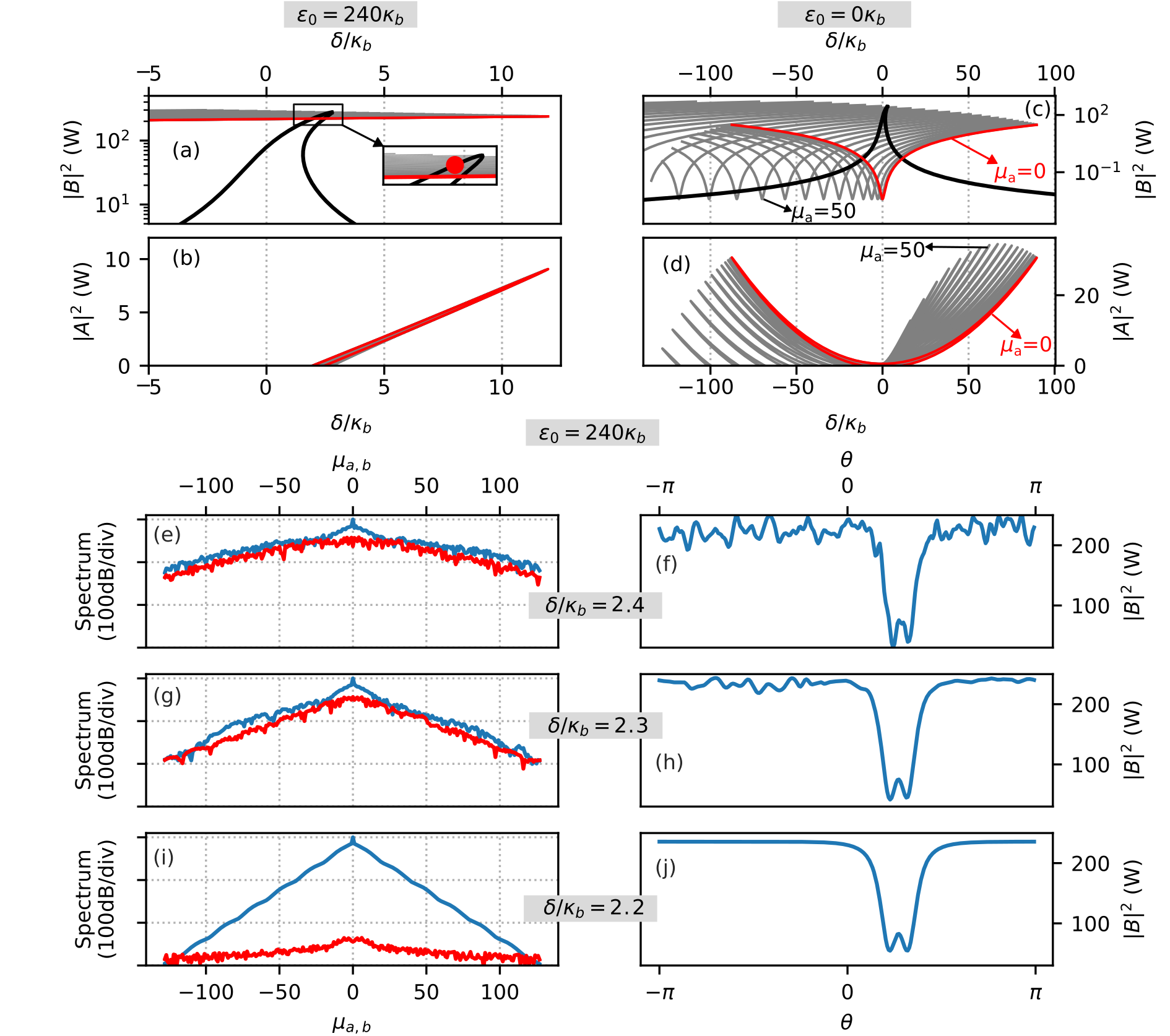}
  \caption{  
  Excitation of a stable dark soliton with  $\mathcal{W}=750$ mW, $\varepsilon_0=240 \kappa_b$, $\kappa_a=\kappa_b=500$MHz. (a, c) Pump and (b,d) half-harmonic intracavity powers for $\varepsilon_0=240$ and $\varepsilon_0=0$, respectively. Gray and red lines represent the analytic three-mode approximations for the OPO ($\mu_a=\pm m$, with $m>0$) and dOPO  ($\mu_a=0$) systems.
(e,g,i) Spectra and (f,h,j) spatial intracavity pump field profiles, respectively, for different detunings. }
  \label{Fig4:DKS}
\end{figure*}

In the cases shown in Figs. \ref{Fig2:stair_OPO} and \ref{Fig3:stair2}, the pump resonance in the absence of the OPO field deviates only weakly from the corresponding linear resonance, indicating that the Kerr nonlinearity has a negligible effect on the pump dynamics. However, increasing the pump power into the sub-kilowatt range, which remains entirely practical experimentally, gives rise to a strong Kerr response and induces bistability of the pump already in the absence of the OPO field, see Fig.~\ref{Fig4:DKS}(a). The resulting intense intracavity field on the upper branch of the bistability loop is not susceptible to Kerr-induced modulational instability because the pump operates in the normal-dispersion regime. Therefore, if the OPO interaction is neglected, the system is expected to support Kerr dark solitons and their associated frequency combs under these conditions~\cite{dark1,dark2,dark3}.

Parametric instabilities, on the other hand, are highly sensitive to both the magnitude and sign of the phase matching parameter, $\varepsilon_0$, and are less critically dependent on the dispersion sign, which therefore provides an effective and distinct means of controlling them. In particular, we find that sufficiently large positive values of $\varepsilon_0$ strongly suppress parametric instabilities. This behaviour is expected because Eq.~\ref{da} implies that for large positive $\varepsilon_0$ the HH mode is necessarily far detuned from resonance, while the pump detuning $\delta_{0b}$ can still be maintained close to resonance.

The example shown in Fig.~\ref{Fig4:DKS}(a,b) corresponds to $\varepsilon_0=240\kappa_b$, which is approximately equivalent to the phase-matching condition violated by only two cavity resonances, while a resonator without a periodic poling would be mismatched by $140$ resonances, see Eq. \ref{eps1}. Consequently, the resonator remains close enough to phase matching to sustain the OPO interaction, while the parametric instabilities occur only in the narrow range of detunings near the tip of the bistable resonance. Shifting $\varepsilon_0$ towards zero and further to the negative values makes the entire upper branch of the bistability loops subject to parametric instabilities (see Fig.~\ref{Fig4:DKS}(c,d)).

The stable homogeneous pump state generally acts as a strong dynamical attractor. Consequently, the excitation of Kerr dark solitons by scanning through the stable continuous-wave state is not possible. Instead, specialised excitation schemes must be employed, including device architectures based on coupled resonators and mode crossing \cite{dark1,dark2}. By contrast, as we demonstrate here, controlling OPO instabilities through phase-mismatch engineering enables straightforward excitation of dark solitons by frequency scanning along a single-mode resonance, thereby providing a considerably simpler route to accessing this regime.

If OPO instabilities arise on the left-hand side of the bistability loop, i.e., at negative detunings, then the continuous-wave state is destabilised before reaching the energy level required to sustain a dark soliton. On the other hand, when OPO generation occurs at high power levels and large detunings, as illustrated in Figs.~\ref{Fig4:DKS}(a,b), the down-conversion process can act as a controlled destabilising mechanism, with pump depletion triggering the transition towards a dark-soliton state.

Figure~\ref{Fig4:DKS}(b) shows that the down-converted powers reach values of up to $\sim 10$ W  only at sufficiently large detunings. At the operating point considered here, the average HH power remains at the level of $\sim 1$--$2$ W. Although significantly weaker than the pump (with a homogeneous state power level $\sim 230$ W
), the HH field is nevertheless sufficiently strong to induce noticeable pump depletion, thereby perturbing the continuous-wave pump state and triggering a dark soliton. 

To excite a dark soliton, we start our frequency scan for low enough detunings, ensuring stability of the CW state and move right towards the narrow range of parametric instabilities. After the instabilities kick in, the dark soliton structure in the pump field is excited on a noisy background (see Figs. \ref{Fig4:DKS}(e-f)). Then we start ramping the frequency back (Figs. \ref{Fig4:DKS}(g-h)), exit the instability range, and end up with a noise-free dark soliton (Figs. \ref{Fig4:DKS}(i-j)). Figure 4 illustrates this approach, while a movie in the supplementary material visualises full dynamics. The proposed technique is of paradigmatic importance for future research into frequency comb generation across the visible spectral range, where dispersion is typically normal and robust techniques for observing frequency combs are currently scarce~\cite{lonc}.

 \section{Bright solitons}
Considering $\varepsilon_0$ and $\delta_{0a}$ as independent parameters, see Eqs.~\ref{eps1} and \ref{db}, and assuming sufficiently large phase mismatch, $|\varepsilon_0|\gg |\delta_{0a}|$, the pump field, i.e., the $b_\mu$ modes, becomes substantially more detuned from resonance than the HH field. In this regime, the pump can be regarded as adiabatically following the dynamics of the HH field.

Indeed, neglecting the pump dispersion, one finds
$b_0 \simeq -\frac{\gamma_{2b}}{\varepsilon_0}a_0^2$ in the leading order.
Substituting this expression into the quadratic coupling term $\gamma_{2a}b_0a_0^*$ in the $a_0$ equation yields an effective Kerr-like contribution proportional to 
$-\frac{\gamma_{2a}\gamma_{2b}}{\varepsilon_0}|a_0|^2a_0$.
Therefore, a positive effective Kerr nonlinearity, capable of compensating anomalous dispersion, is obtained for $\varepsilon_0<0$.

The system thus approaches a limit in which the HH field is qualitatively described by a parametrically driven Ginzburg--Landau equation with anomalous dispersion and positive Kerr nonlinearity, suggesting the existence of bright parametric solitons \cite{opex,param}. Within this picture, the dispersion sign of the pump field becomes of secondary importance, since the pump pulse is predominantly shaped by nonlinear driving terms with dominant HH contributions, such as $A^2$, rather than by its intrinsic dispersive and nonlinear properties. Consistent with this viewpoint, Ref.~\cite{bruch2021} reported the observation of bright solitons in the HH field accompanied by a coherent pump waveform that did not exhibit a bright-soliton profile, despite both fields experiencing anomalous dispersion. A further interpretation of these results was subsequently presented in Ref.~\cite{ding2024}.

\begin{figure*}[!h]
\includegraphics[width=\textwidth]{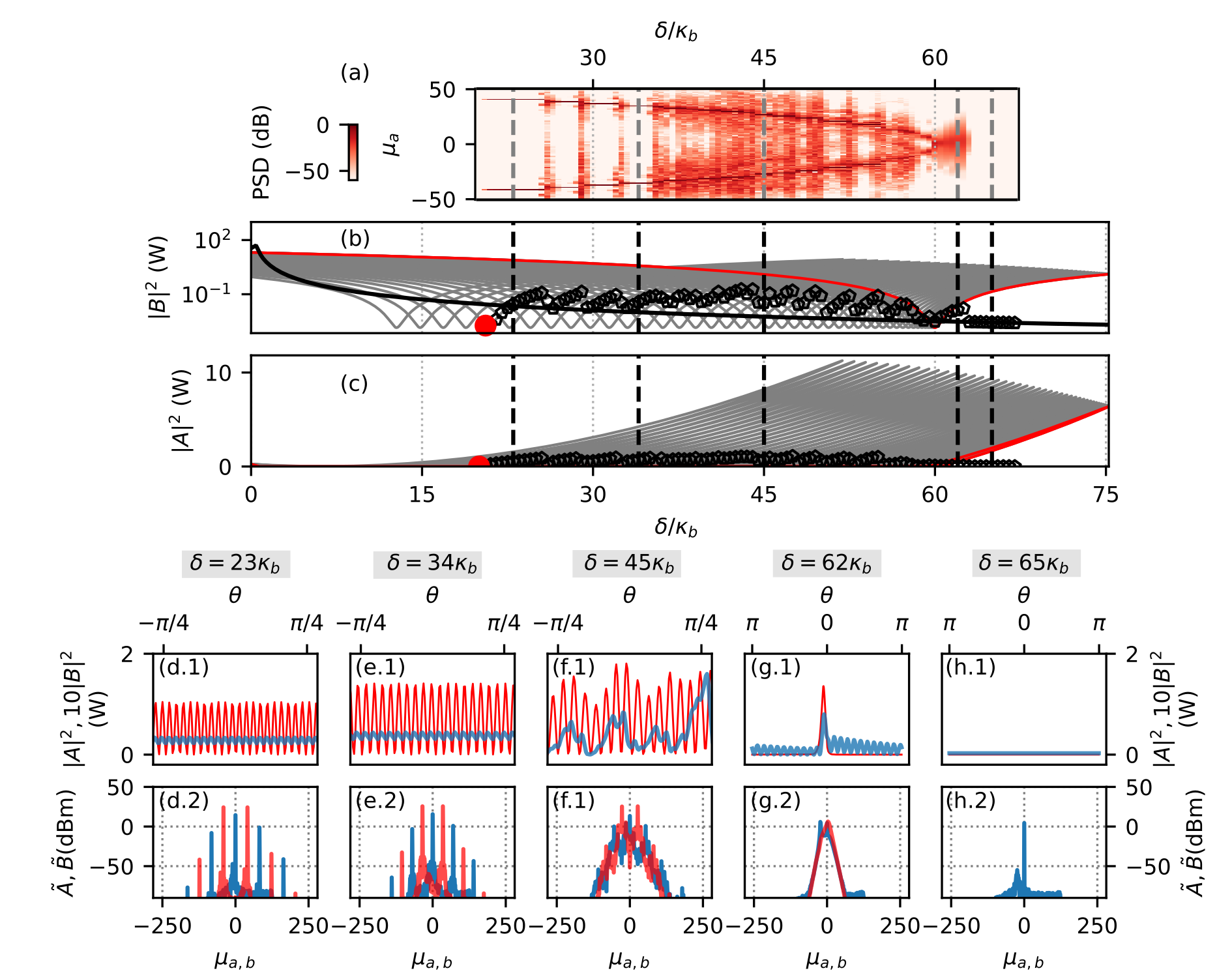}
  \caption{ (a, b, c) Parametric instability structures for $\mathcal{W}=150$ mW and $\varepsilon_0=-60 \kappa_b$.(d,f,h,i) three-breathers state: (d) amplitudes, (f) real parts, (h) breathing dynamics and (i) spectra. (e,g,j) single bright-to-bright cavity soliton state: (e) amplitudes, (g) real parts and (j) spectra.}
  \label{Fig5:bright_sol}
\end{figure*}
We note that the above observations are distinct from the two-colour bright OPO solitons predicted when both field components simultaneously exhibit well-developed canonical bright-soliton profiles~\cite{dmitry1,pedro2,leo2}. Such canonical states have not yet been observed experimentally and, to the best of our knowledge, have not been found in realistic microresonator models incorporating typical experimental dispersion profiles and thereby represent one of the outstanding challenges.

A necessary condition for the existence of bright solitons in the parametrically driven Ginzburg--Landau model is the coexistence of a stable zero-amplitude (non-OPO) background and a modulationally unstable dOPO state. To reveal the anticipated bright-soliton excitations in our system, we set $\varepsilon_0=-60\kappa_b$. This intermediate value allows the above asymptotic picture to serve as a useful guideline while keeping the required pump powers at moderate levels, since the effective Kerr nonlinearity scales as $|\varepsilon_0|^{-1}$. 

For detuning values below the soliton-existence interval, higher pump powers drive the system from a clearly three-mode OPO configuration (such as in Fig.~\ref{Fig2:stair_OPO}(a)) into a more chaotic state with spectral components still peaked around a non-degenerate OPO discrete staircase evolution (see Fig.~\ref{Fig5:bright_sol}(a)).
For larger detuning, the coexistence of the non-OPO and dOPO states is observed for $60\lesssim\delta/\kappa_b\lesssim 75$, or equivalently $23\lesssim\delta_{0a}/\kappa_b\lesssim 30$, see Figs.~\ref{Fig5:bright_sol}(b-c). Within this interval, two-colour bright soliton states are readily excited.

As anticipated from the asymptotic arguments above, the HH component dominates the overall waveform, while the pump pulse remains substantially weaker.
The pump pulse frequently develops pronounced oscillatory tails, making its classification as a soliton ambiguous in the traditional sense \cite{Bengel2026OL, tal25}. Nevertheless, its overall coherence supports a broader interpretation of the two-field structure as a composite two-colour soliton state.
We report various stages of OFC generation, from patterns formation (Figs. \ref{Fig5:bright_sol}(d.1,2) and (e.1,2)), to chaos (Figs. \ref{Fig5:bright_sol}(f.1,2)), solitons (Figs. \ref{Fig5:bright_sol}(g.1,2)) and non-OPO state (Figs. \ref{Fig5:bright_sol}(h.1,2)).

\section{Topological solitons}
Topological solitons in parametric systems correspond to phase trajectories connecting two fixed points associated with opposite phases of the signal field, as implied by the symmetry transformation: 
\begin{equation}
(a_{\mu},b_{\mu})\to (-a_{\mu},b_{\mu}).
\label{refl}
\end{equation}
In a ring geometry, such states can only exist in pairs to satisfy the periodic boundary conditions. Topological states in parametric systems have been extensively investigated across a variety of models and physical configurations, as documented in several foundational works ~\cite{pedro2019,topo,oppo99,oppo01,old1}. However, their appearance in high-repetition-rate microresonators, where the coupled-mode description appropriately captures the dynamics ~\cite{bruch2021,lu2023,josab}, has not yet been investigated.

\begin{figure*}[!t]
\includegraphics[width=\textwidth]{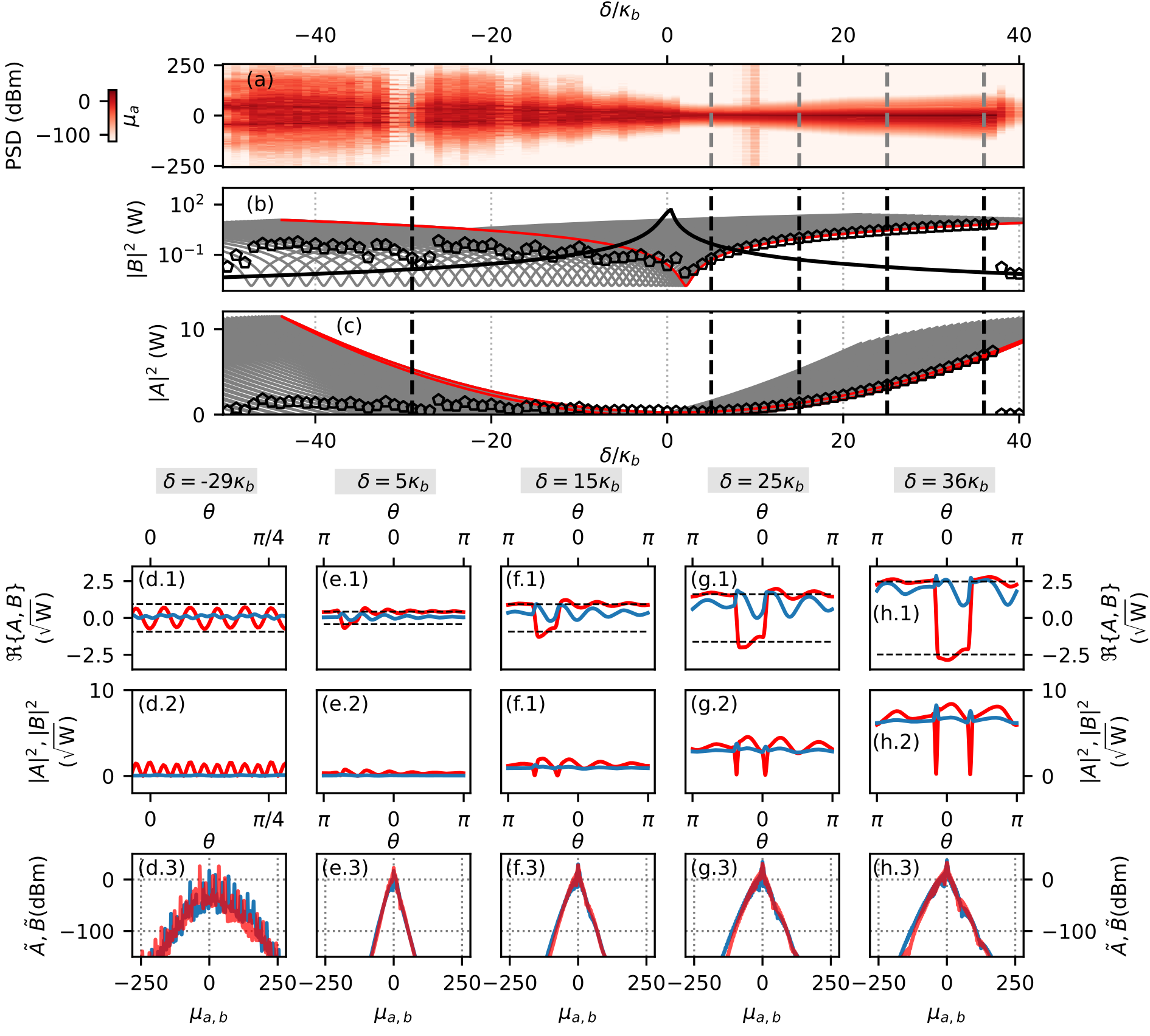}
  \caption{(a, b, c) Parametric instability structures for ${\cal W} = 150 $ mW and $\varepsilon_0=-2.1 \kappa_b$. Vertical dashed lines locate specific solutions whose spatial and spectral profiles are reported on panels (d-h). The gray dashed horizontal lines mark the dOPO solutions considering a $\pi$-phase factor degeneracy for the HH field. }
  \label{Fig6:TOPO}
\end{figure*}

We find that for relatively small values of $\varepsilon_0$ the dOPO state possesses a broad parameter range in which it remains stable against modulational instabilities. This stability is a prerequisite for the existence of topological solitons connecting the symmetry-related dOPO states with opposite phases, see Eq.\ref{refl}. 
The dOPO and nondegenerate OPO states corresponding to the selected value of $\varepsilon_0$ are shown in Figs.~\ref{Fig6:TOPO}(a-c). Here, we present the spectral evolution of field $A$ as a function of $\delta$, along with the intracavity field amplitudes and parametric instabilities, following the same procedure as in the previous cases.

\begin{figure*}[t]
\includegraphics[width=\textwidth]{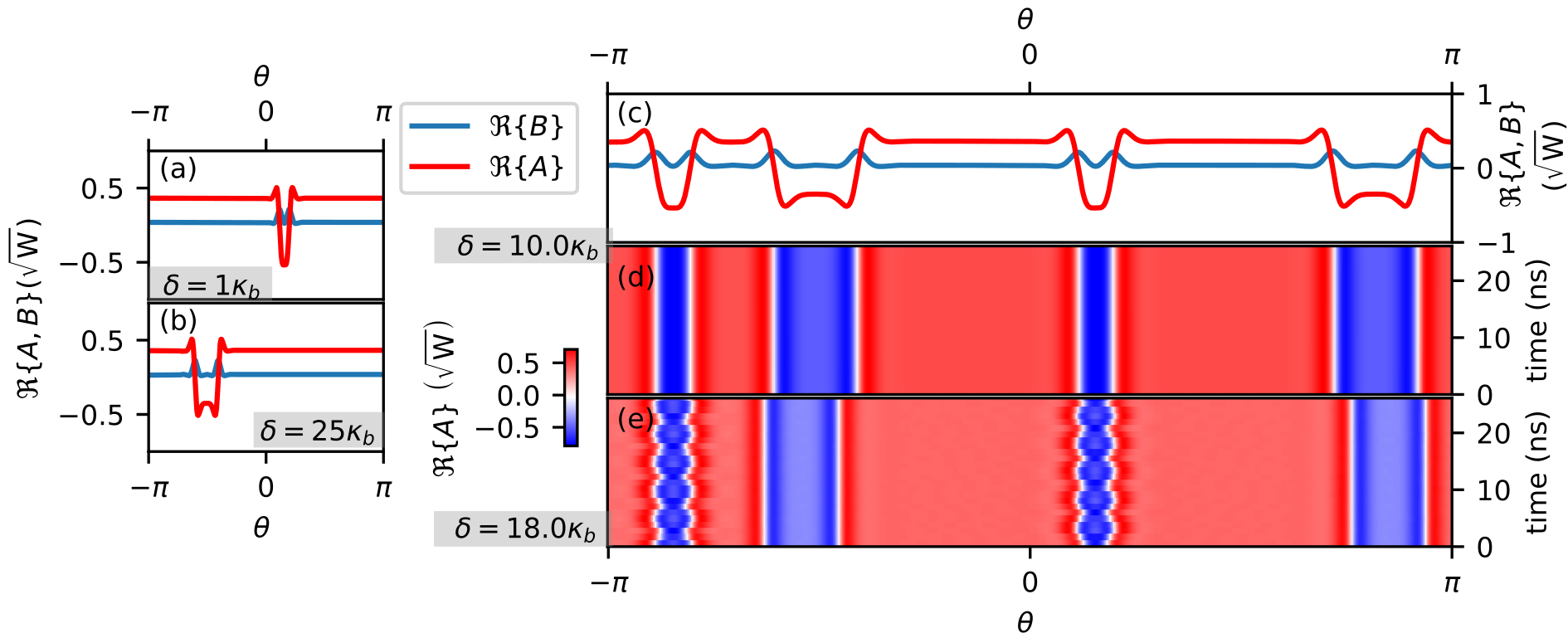}
  \caption{Topological solitons in the zero walk-off limit: (a, b) two  solitons excited at different detunings. (c) Spatial profiles of the coexisting topological solitons and (d) space-time traces showing their stability for a chosen detuning. (e) Coexistence of the breathing and static topological solitons for  larger detuning. }
  \label{Fig7:bistable}
\end{figure*}
To highlight the topological signature of the system, the real part of the fields is plotted alongside the dOPO solutions (grey dashed lines) in the top row of the selected localised states, corresponding to the coordinates of the vertical dashed lines in panels (a–c). As observed in Fig.~\ref{Fig6:TOPO}(d.1-3), the modulation instability regimes are governed by a field oscillation comprised within the dOPO solutions and characterised by a $\pi$-phase jump. Interestingly, as $\delta$ increases toward a red-detuned configuration, a localised state becomes locked around the phase jump (Fig.~\ref{Fig6:TOPO}(e.1-3)) and its amplitude increases at larger detunings (Fig.~\ref{Fig6:TOPO}(f-h)).

Although the system clearly underpins a phase jump of the intracavity field $A$ that locks a localised state, the background level of the dOPO solution is not perfectly flat; instead, it exhibits small oscillations. These features arise from a non-negligible group velocity, which obscures the otherwise clear topological signature. The resulting cavity soliton state is nonetheless a robust dynamical attractor, stable against noise and external perturbations, and characterised by a wide locking range.

It is instructive to show some characteristic solutions in the limit of negligible walk-off, i.e. $D_{1a}\rightarrow D_{1b}$. This scenario could become realistic if, e.g., different families of modes are excited in the pump and half-harmonic fields \cite{luo2019modalphsematching,ciret2020,gerini2025,tal25}. In Fig.\ref{Fig7:bistable} (a,b), we show two distinct topological solitons excited at different detunings ($\delta=1$ and $25$ times $\kappa_b$) with different core sizes. We observe that the background flatness is significantly enhanced compared to the walk-off case. Notably, these two distinct solutions coexist within a finite range of laser-cavity detuning, see Fig.~\ref{Fig7:bistable}(c). At lower detunings ($\delta=10 \kappa_b$), both solutions remain stable (Fig.~\ref{Fig7:bistable}(d)). However, as $\delta$ increases ($\delta=10 \kappa_b$), the narrower core soliton loses stability and begins to oscillate, see Fig.~\ref{Fig7:bistable}(e) and Supplemental dynamical visualisations.

\newpage


\section{Conclusions \label{sec:conclusion}}
\begin{figure*}[!b]
\includegraphics[width=\textwidth]{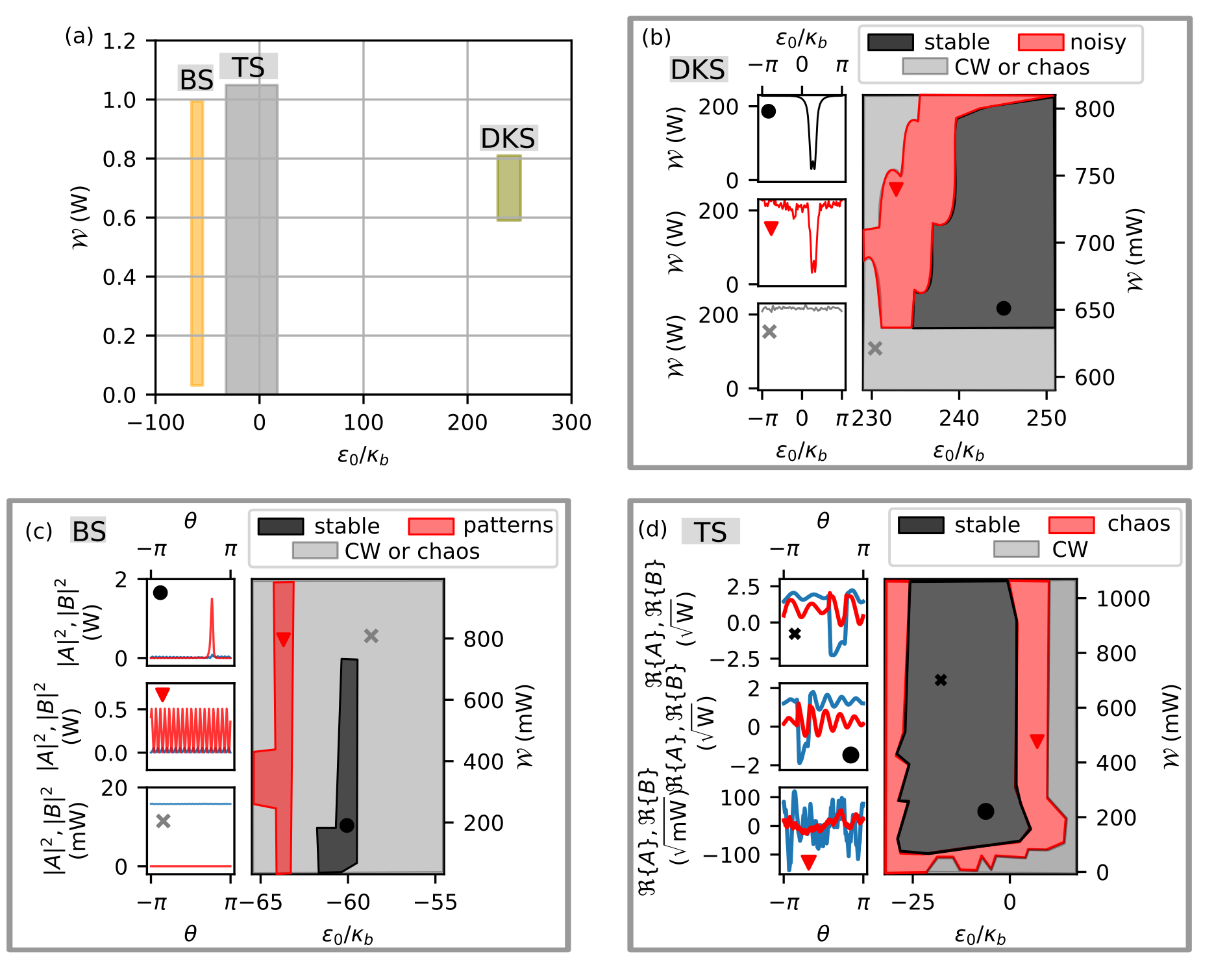}
  \caption{(a) The $\delta$ versus $\cal W$ phase diagram mapping the various localized states discussed in this letter. (b) Dark solitons (DKS) are classified as stable, noisy, chaotic, or collapsed to the continuous-wave (CW) state. (c) For bright solitons (BS), we highlight patterns and stable solutions; (d) for topological solitons, we mark stable states and chaotic or CW solutions. The scatters locate the coordinates of the specific solutions reported in the insets.
}
  \label{Fig9:phase_diagram}
\end{figure*}
We considered a technologically relevant lithium-niobate microresonator geometry with a high repetition rate and modelled it using coupled-mode equations. The system is pumped at 775 nm and phase-matched for half-harmonic generation at 1550 nm. Experimental studies of such setups have so far been limited to the regime involving a single pump mode and a pair of half-harmonic sidebands~\cite{lu21}.

Our modelling shows that, by controlling phase matching through periodic poling and temperature tuning, this platform can support a wide variety of nonlinear states, including bright, dark, and topological solitons. These results substantially expand the expected operational regimes of half-harmonic microresonators and motivate further experimental investigations.

It is instructive to map the presented $\chi^{(2)}+\chi^{(3)}$ soliton regimes onto the pump power and phase-matching parameter space ($\varepsilon_0$ vs $\cal W$) to illustrate the vast operational range opened for further research. As illustrated in Fig.~\ref{Fig9:phase_diagram}(a), the various localised states investigated in this work span widely separated regions of the phase space. Specifically, topological solitons reside at relatively low input pump powers and tight phase-matching configurations. In this regime, the minimal-power OPO trigger points are aligned with the cavity resonances. Hence, the dynamics is governed by $\chi^{(2)}$ effects, since Kerr thresholds are situated at much higher values of power. Conversely, both bright and dark solitons require highly phase-mismatched configurations to shift the parametric instability boundaries into power regimes suitable for their excitation, following the prescriptions established herein. The bright soliton case represents a regime where both $\chi^{(2)}$ and $\chi^{(
3)}$ effects are simultaneously important, whereas $\chi^{(
3)}$ effects predominantly govern the dark-soliton regime.

Finally, we explore the robustness of these localised states by mapping their stability domains over the phase-matching parameter space ($\varepsilon_0$ versus $\cal W$), as reported in Figs.~\ref{Fig9:phase_diagram}(b,c,d).
Importantly, all soliton states reported here, including the dark-soliton and topological-soliton regimes, were generated using a simple pump-frequency scan, suggesting a comparatively straightforward experimental pathway for their realisation.

\section{Funding}
This work was supported by the UK Engineering and Physical Sciences Research Council (Grant No. EP/X040844/1) and  the Royal Society 
(Grant No. IF/R2/252184). NPL acknowledges the support of the UK Government Department for Science, Innovation and Technology through the UK National Quantum Technologies Programme.
\section{Data Availability Statement}
The data supporting the findings of this study are publicly available.


\begin{thebibliography}{0}%
\makeatletter
\providecommand \@ifxundefined [1]{%
 \@ifx{#1\undefined}
}%
\providecommand \@ifnum [1]{%
 \ifnum #1\expandafter \@firstoftwo
 \else \expandafter \@secondoftwo
 \fi
}%
\providecommand \@ifx [1]{%
 \ifx #1\expandafter \@firstoftwo
 \else \expandafter \@secondoftwo
 \fi
}%
\providecommand \natexlab [1]{#1}%
\providecommand \enquote  [1]{``#1''}%
\providecommand \bibnamefont  [1]{#1}%
\providecommand \bibfnamefont [1]{#1}%
\providecommand \citenamefont [1]{#1}%
\providecommand \href@noop [0]{\@secondoftwo}%
\providecommand \href [0]{\begingroup \@sanitize@url \@href}%
\providecommand \@href[1]{\@@startlink{#1}\@@href}%
\providecommand \@@href[1]{\endgroup#1\@@endlink}%
\providecommand \@sanitize@url [0]{\catcode `\\12\catcode `\$12\catcode `\&12\catcode `\#12\catcode `\^12\catcode `\_12\catcode `\%12\relax}%
\providecommand \@@startlink[1]{}%
\providecommand \@@endlink[0]{}%
\providecommand \url  [0]{\begingroup\@sanitize@url \@url }%
\providecommand \@url [1]{\endgroup\@href {#1}{\urlprefix }}%
\providecommand \urlprefix  [0]{URL }%
\providecommand \Eprint [0]{\href }%
\providecommand \doibase [0]{https://doi.org/}%
\providecommand \selectlanguage [0]{\@gobble}%
\providecommand \bibinfo  [0]{\@secondoftwo}%
\providecommand \bibfield  [0]{\@secondoftwo}%
\providecommand \translation [1]{[#1]}%
\providecommand \BibitemOpen [0]{}%
\providecommand \bibitemStop [0]{}%
\providecommand \bibitemNoStop [0]{.\EOS\space}%
\providecommand \EOS [0]{\spacefactor3000\relax}%
\providecommand \BibitemShut  [1]{\csname bibitem#1\endcsname}%
\let\auto@bib@innerbib\@empty
\end{thebibliography}%


\newpage
\begin{thebibliography}{10}

\bibitem{didrev}
Scott~A Diddams, Kerry Vahala, and Thomas Udem.
\newblock Optical frequency combs: Coherently uniting the electromagnetic spectrum.
\newblock {\em Science}, 369(6501):eaay3676, 2020.

\bibitem{Bellini:1998}
M~Bellini, C~Lynga, A~Tozzi, MB~Gaarde, TW~H{\"a}nsch, and C-G Wahlstr{\"o}m.
\newblock Temporal coherence of ultrashort high-order harmonic pulses.
\newblock {\em Physical review letters}, 81(2):297, 1998.

\bibitem{Diddams:2000}
Scott~A. Diddams, David~J. Jones, Jun Ye, Steven~T. Cundiff, John~L. Hall, Jinendra~K. Ranka, Robert~S. Windeler, Ronald Holzwarth, Thomas Udem, and T.~W. H\"ansch.
\newblock Direct link between microwave and optical frequencies with a 300 {THz} femtosecond laser comb.
\newblock {\em Phys. Rev. Lett.}, 84:5102--5105, May 2000.

\bibitem{ulvila2013}
Ville Ulvila, Christopher~R Phillips, Lauri Halonen, and Markku Vainio.
\newblock Frequency comb generation by a continuous-wave-pumped optical parametric oscillator based on cascading quadratic nonlinearities.
\newblock {\em Optics letters}, 38(21):4281--4284, 2013.

\bibitem{Ricciardi2015}
Iolanda Ricciardi, Simona Mosca, Maria Parisi, Pasquale Maddaloni, Luigi Santamaria, Paolo De~Natale, and Maurizio De~Rosa.
\newblock Frequency comb generation in quadratic nonlinear media.
\newblock {\em Physical Review A}, 91(6):063839, 2015.

\bibitem{Mosca2018}
S.~Mosca, M.~Parisi, I.~Ricciardi, F.~Leo, T.~Hansson, M.~Erkintalo, P.~Maddaloni, P.~De~Natale, S.~Wabnitz, and M.~De~Rosa.
\newblock Modulation instability induced frequency comb generation in a continuously pumped optical parametric oscillator.
\newblock {\em Phys. Rev. Lett.}, 121:093903, Aug 2018.

\bibitem{iolandarev}
I.~Ricciardi, S.~Mosca, M.~Parisi, F.~Leo, T.~Hansson, M.~Erkintalo, P.~Maddaloni, P.~De~Natale, S.~Wabnitz, and M.~De~Rosa.
\newblock Optical frequency combs in quadratically nonlinear resonators.
\newblock {\em Micromachines}, 11:230, 2020.

\bibitem{jank}
M.~Jankowski, J.~Mishra, and M.~M. Fejer.
\newblock Dispersion-engineered $\chi^{(2)}$ nanophotonics: a flexible tool for nonclassical light.
\newblock {\em J. Phys. Photon.}, 3:042005, 2021.

\bibitem{boes}
Andreas Boes, Lin Chang, Carsten Langrock, Mengjie Yu, Mian Zhang, Qiang Lin, Marko Lon\v{c}ar, Martin Fejer, John Bowers, and Arnan Mitchell.
\newblock Lithium niobate photonics: Unlocking the electromagnetic spectrum.
\newblock {\em Science}, 379(6627):eabj4396, 2023.

\bibitem{Savchenkov:2004}
Anatoliy~A. Savchenkov, Andrey~B. Matsko, Dmitry Strekalov, Makan Mohageg, Vladimir~S. Ilchenko, and Lute Maleki.
\newblock Low threshold optical oscillations in a whispering gallery mode ${\mathrm{c}\mathrm{a}\mathrm{f}}_{2}$ resonator.
\newblock {\em Phys. Rev. Lett.}, 93:243905, Dec 2004.

\bibitem{Kippenberg:2004}
T.~J. Kippenberg, S.~M. Spillane, and K.~J. Vahala.
\newblock Kerr-nonlinearity optical parametric oscillation in an ultrahigh-{Q} toroid microcavity.
\newblock {\em Phys. Rev. Lett.}, 93:083904, Aug 2004.

\bibitem{Haye:2007Nat}
Pascal Del'Haye, Albert Schliesser, Olivier Arcizet, Tom Wilken, Ronald Holzwarth, and Tobias~J Kippenberg.
\newblock Optical frequency comb generation from a monolithic microresonator.
\newblock {\em Nature}, 450(7173):1214--1217, 2007.

\bibitem{kiprev}
T.~J. Kippenberg, A.~L. Gaeta, M.~Lipson, and M.~L. Gorodetsky.
\newblock Dissipative kerr solitons in optical microresonators.
\newblock {\em Science}, 361:eaan8083, 2018.

\bibitem{Herr:2014NATPHOT}
Tobias Herr, Victor Brasch, John~D Jost, Christine~Y Wang, Nikita~M Kondratiev, Michael~L Gorodetsky, and Tobias~J Kippenberg.
\newblock Temporal solitons in optical microresonators.
\newblock {\em Nature Photonics}, 8(2):145--152, 2014.

\bibitem{Haye:2011PRL}
P.~Del'Haye, T.~Herr, E.~Gavartin, M.~L. Gorodetsky, R.~Holzwarth, and T.~J. Kippenberg.
\newblock Octave spanning tunable frequency comb from a microresonator.
\newblock {\em Phys. Rev. Lett.}, 107:063901, Aug 2011.

\bibitem{nie}
M.~Nie, Y.~Xie, B.~Li, and S.-W. Huang.
\newblock Photonic frequency microcombs based on dissipative kerr and quadratic cavity solitons.
\newblock {\em Progress in Quantum Electronics}, 86:100437, 2022.

\bibitem{skrrev}
N.~Englebert, R.~M. Gray, A.~Marandi, Markku Vainio, Zheng Gong, Hong~X. Tang, and Dmitry~V. Skryabin.
\newblock Temporal solitons and frequency combs in quadratic resonators.
\newblock {\em Nat. Photon.}, 20:616--627, 2026.

\bibitem{tal25}
F.~R. Talenti, L.~Lovisolo, A.~Gerini, H.~Peng, P.~Parra-Rivas, T.~Hansson, Y.~Sun, C.~Alonso-Ramos, M.~Morassi, A.~Lema{\^\i}tre, et~al.
\newblock Interplay of $\chi^{(2)}$ and $\chi^{(3)}$ effects for microcomb generation.
\newblock {\em J. Eur. Opt. Soc.-Rapid Publ.}, 21(1):23, 2025.

\bibitem{moille2018OL}
Gregory Moille, Qing Li, Sangsik Kim, Daron Westly, and Kartik Srinivasan.
\newblock Phased-locked two-color single soliton microcombs in dispersion-engineered {Si3N4} resonators.
\newblock {\em Optics Letters}, 43(12):2772--2775, 2018.

\bibitem{moille2023fourier}
Gr{\'e}gory Moille, Xiyuan Lu, Jordan Stone, Daron Westly, and Kartik Srinivasan.
\newblock Fourier synthesis dispersion engineering of photonic crystal microrings for broadband frequency combs.
\newblock {\em Communications Physics}, 6(1):144, 2023.

\bibitem{Tal22}
Francesco~Rinaldo Talenti, Stefan Wabnitz, In\`es Ghorbel, Sylvain Combri\'e, Luca Aimone-Giggio, and Alfredo De~Rossi.
\newblock Fast dispersion tailoring of multimode photonic crystal resonators.
\newblock {\em Phys. Rev. A}, 106:023505, Aug 2022.

\bibitem{lucas2023tailoring}
Erwan Lucas, Su-Peng Yu, Travis~C Briles, David~R Carlson, and Scott~B Papp.
\newblock Tailoring microcombs with inverse-designed, meta-dispersion microresonators.
\newblock {\em Nature Photonics}, 17(11):943--950, 2023.

\bibitem{tal26}
Francesco~Rinaldo Talenti, Luca Lovisolo, Zijun Xiao, Zeina Saleh, Andrea Gerini, Carlos Alonso-Ramos, Martina Morassi, Aristide Lema{\^\i}tre, Abdelmounaim Harouri, Stefan Wabnitz, et~al.
\newblock Dispersion engineered algaas-on-insulator nanophotonics by distributed feedback.
\newblock {\em ACS Photonics}, 13(3):774--781, 2026.

\bibitem{he2019}
Yang He, Qi-Fan Yang, Jingwei Ling, Rui Luo, Hanxiao Liang, Mingxiao Li, Boqiang Shen, Heming Wang, Kerry Vahala, and Qiang Lin.
\newblock Self-starting bi-chromatic {LiNbO}3 soliton microcomb.
\newblock {\em Optica}, 6(9):1138--1144, 2019.

\bibitem{vil19}
A.~Villois, N.~Kondratiev, I.~Breunig, D.~N. Puzyrev, and D.~S. Skryabin.
\newblock Frequency combs in a microring optical parametric oscillator.
\newblock {\em Opt. Lett.}, 44(18):4443--4446, 2019.

\bibitem{bruch2021}
A.~W. Bruch, X.~Liu, Z.~Gong, J.~B. Surya, M.~Li, C.-L. Zou, and H.~X. Tang.
\newblock Pockels soliton microcomb.
\newblock {\em Nature Photon.}, 15:21, 2021.

\bibitem{ding2024}
Y.~Ding, Z.~Wei, Y.~Wang, C.~Yang, and C.~Bao.
\newblock Theoretical analysis of microcavity simultons reinforced by $\chi^{(2)}$ and $\chi^{(3)}$ nonlinearities.
\newblock {\em Phys. Rev. Lett.}, 132(1):013801, 2024.

\bibitem{simult}
G.~Wu, Y.~Wei, L.~Li, S.~Chen, L.~Bu, F.~Baronio, T.~Lin, M.~Zhu, S.~Trillo, and Z.~Ni.
\newblock Ultraflat soliton microcombs in driven quadratic-kerr nonlinear microresonators.
\newblock {\em Phys. Rev. Lett.}, 135(11):113801, 2025.

\bibitem{Bengel2026OL}
Lukas Bengel, Francesco~Rinaldo Talenti, Ahmadreza Alaeddini, Carolin Scheib, Bj{\"o}rn~de Rijk, Wolfgang Reichel, Giuseppe Leo, Christian Koos, Huanfa Peng, and Stefan Wabnitz.
\newblock Second-harmonic bichromatic dispersive wave comb generation in a dissipative kerr temporal soliton fabry--perot.
\newblock {\em Optics Letters}, 51(9):2648--2651, 2026.

\bibitem{Sun26}
Yifan Sun, Clément Dupont, Edem Kossi~Akakpo, Francesco De~Lucia, Georges Semaan, Simon-Pierre Gorza, and François Leo.
\newblock {\em Journal of Physics: Photonics}, 8(3):03LT01, 2026.

\bibitem{pedro2019}
P.~Parra-Rivas, L.~Gelens, T.~Hansson, S.~Wabnitz, and F.~Leo.
\newblock Frequency comb generation through the locking of domain walls in doubly resonant dispersive optical parametric ovens.
\newblock {\em Opt. Lett.}, 44(8):2004--2007, 2019.
\bibitem{topo}
N.~Englebert, R.~M. Gray, L.~Ledezma, R.~Sekine, T.~Zacharias, R.~Ramesh, B.~K. Gutierrez, P.~Parra-Rivas, and A.~Marandi.
\newblock Topological soliton frequency comb in nanophotonic lithium niobate.
\newblock {\em Nature}, pages 1--6, 2026.

\bibitem{phil24}
C.~R. Phillips, M.~Jankowski, N.~Flemens, and M.~M. Fejer.
\newblock General framework for ultrafast nonlinear photonics: unifying single and multi-envelope treatments.
\newblock {\em Opt. Express}, 32(5):8284--8307, 2024.

\bibitem{tal25OL}
F.~R. Talenti, S.~Wabnitz, Y.~Sun, T.~Hansson, L.~Lovisolo, A.~Gerini, G.~Leo, L.~Vivien, C.~Koos, H.~Peng, and P.~Parra-Rivas.
\newblock Bistable soliton optical frequency combs in a second-harmonic generation kerr cavity.
\newblock {\em Opt. Lett.}, 50:2037--2040, 2025.

\bibitem{hill2020effects}
Lewis Hill, Gian-Luca Oppo, Michael~TM Woodley, and Pascal Del'Haye.
\newblock Effects of self-and cross-phase modulation on the spontaneous symmetry breaking of light in ring resonators.
\newblock {\em Physical Review A}, 101(1):013823, 2020.

\bibitem{lucas2025faticons}
Erwan Lucas, Gang Xu, Pengxiang Wang, Gian-Luca Oppo, Lewis Hill, Pascal Del'Haye, Bertrand Kibler, Yiqing Xu, Stuart~G Murdoch, Miro Erkintalo, et~al.
\newblock Polarization faticons: Chiral localized structures in self-defocusing kerr resonators.
\newblock {\em Physical Review Letters}, 135(6):063803, 2025.

\bibitem{josab}
D.~V. Skryabin.
\newblock Coupled-mode theory for microresonators with quadratic nonlinearity.
\newblock {\em J. Opt. Soc. Am. B}, 37(9):2604--2614, 2020.

\bibitem{puz2}
Danila~N. Puzyrev and Dmitry~V. Skryabin.
\newblock Carrier-resolved real-field theory of multi-octave frequency combs.
\newblock {\em Optica}, 10:770--773, 2023.

\bibitem{puz22}
D.~N. Puzyrev and D.~V. Skryabin.
\newblock Ladder of eckhaus instabilities and parametric conversion in chi(2) microresonators.
\newblock {\em Commun. Phys.}, 5(1):125, 2022.

\bibitem{ingo1}
N.~Amiune, Z.~Fan, V.~V. Pankratov, D.~N. Puzyrev, D.~V. Skryabin, K.~T. Zawilski, P.~G. Schunemann, and I.~Breunig.
\newblock Mid-infrared frequency combs and staggered spectral patterns in $\protect\chi^{(2)}$ microresonators.
\newblock {\em Opt. Express}, 31:907--915, 2023.

\bibitem{ingo2}
N.~Amiune, D.~N. Puzyrev, V.~V. Pankratov, D.~V. Skryabin, K.~Buse, and I.~Breunig.
\newblock Optical-parametric-oscillation-based $\protect\chi^{(2)}$ frequency comb in a lithium niobate microresonator.
\newblock {\em Opt. Express}, 29:41378--41387, 2021.
\bibitem{luo2019modalphsematching}
Luo R, He Y, Liang H, Li M, Ling J and Lin Q 2019 \textit{Physical Review Applied} \textbf{11} 034026

\bibitem{yang2024}
Yang F, Lu J, Shen M, Yang G and Tang H X 2024 \textit{Optica} \textbf{11} 1050--1055

\bibitem{wolf2018quasi}
Wolf R, Jia Y, Bonaus S, Werner C S, Herr S J, Breunig I, Buse K and Zappe H 2018 \textit{Optica} \textbf{5} 872--875

\bibitem{lu2019}
Lu J, Surya J B, Liu X, Bruch A W, Gong  Z, Xu Y and Tang H X 2019 \textit{Optica} \textbf{6} 1455--1460

\newblock {\em Nature Communications}, 2026.

\bibitem{lonc}
Y.~Song, Z.~Li, X.~Zhu, N.~Lippok, M.~Erkintalo, and M.~Lon\v{c}ar.
\newblock High-efficiency and broadband kerr comb generation in normal-dispersion x-cut lithium niobate microresonators.
\newblock {\em Sci. Adv.}, 12:eaeb5758, 2026.

\bibitem{oppo99}
G.-L. Oppo, A.~J. Scroggie, and W.~J. Firth.
\newblock From domain walls to localized structures in degenerate optical parametric oscillators.
\newblock {\em J. Opt. B: Quantum Semiclass. Opt.}, 1(1):133, 1999.

\bibitem{oppo01}
G.-L. Oppo, A.~J. Scroggie, and W.~J. Firth.
\newblock Characterization, dynamics and stabilization of diffractive domain walls and dark ring cavity solitons in parametric oscillators.
\newblock {\em Phys. Rev. E}, 63:066209, 2001.

\bibitem{old1}
D.~V. Skryabin, A.~Yulin, D.~Michaelis, W.~J. Firth, G.-L. Oppo, U.~Peschel, and F.~Lederer.
\newblock Perturbation theory for domain walls in the parametric ginzburg-landau equation.
\newblock {\em Phys. Rev. E}, 64:056618, 2001.

\bibitem{dark1}
X.~Xue, Y.~Xuan, Y.~Liu, et~al.
\newblock Mode-locked dark pulse kerr combs in normal-dispersion microresonators.
\newblock {\em Nature Photon}, 9:594--600, 2015.

\bibitem{dark2}
I.~Rebolledo-Salgado, C.~Quevedo-Gal{\protect\'a}n, {\protect\'O}.~B. Helgason, et~al.
\newblock Platicon dynamics in photonic molecules.
\newblock {\em Commun Phys}, 6:303, 2023.

\bibitem{dark3}
G.~N. Campbell, L.~Hill, P.~Del'Haye, and G.-L. Oppo.
\newblock Dark solitons in fabry-p{\'e}rot resonators with kerr media and normal dispersion.
\newblock {\em Physical Review A}, 108:033505, 2023.

\bibitem{Zhu2021LN}
Di~Zhu, Linbo Shao, Mengjie Yu, Rebecca Cheng, Boris Desiatov, C.~J. Xin, Yaowen Hu, Jeffrey Holzgrafe, Soumya Ghosh, Amirhassan Shams-Ansari, Eric Puma, Neil Sinclair, Christian Reimer, Mian Zhang, and Marko Lon\v{c}ar.
\newblock Integrated photonics on thin-film lithium niobate.
\newblock {\em Adv. Opt. Photon.}, 13(2):242--352, Jun 2021.

\bibitem{lu2023}
J.~Lu, D.~N. Puzyrev, V.~V. Pankratov, D.~S. Skryabin, F.~Yang, Z.~Gong, J.~B. Surya, and H.~X. Tang.
\newblock Two-colour dissipative solitons and breathers in microresonator second-harmonic generation.
\newblock {\em Nature Commun.}, 14:2798, 2023.

\bibitem{opex}
Dmitry~V. Skryabin.
\newblock Sech-squared pockels solitons in the microresonator parametric down-conversion.
\newblock {\em Opt. Express}, 29:28521--28529, 2021.

\bibitem{param}
N.~Englebert, F.~De~Lucia, P.~Parra-Rivas, et~al.
\newblock Parametrically driven kerr cavity solitons.
\newblock {\em Nat. Photon.}, 15:857--861, 2021.

\bibitem{dmitry1}
D.~V. Skryabin.
\newblock Instabilities of cavity solitons in optical parametric oscillators.
\newblock {\em Physical Review E}, 60(4):R3508, 1999.

\bibitem{pedro2}
P.~Parra-Rivas, et~al.
\newblock Frequency combs and localized states in optical parametric oscillators.
\newblock {\em Physical Review A}, 99:043813, 2019.

\bibitem{leo2}
P.~Parra-Rivas, C.~Mas~Arab{\'\i}, and F.~Leo.
\newblock Dissipative localized states and breathers in phase-mismatched singly resonant optical parametric oscillators: Bifurcation structure and stability.
\newblock {\em Phys. Rev. Research}, 4:013044, 2022.

\bibitem{Kaplan1985}
A.~E. Kaplan.
\newblock Bistable solitons.
\newblock {\em Phys. Rev. Lett.}, 55:1291--1294, Sep 1985.

\bibitem{Derossi1998}
Alfredo De~Rossi, Claudio Conti, and Stefano Trillo.
\newblock Stability, multistability, and wobbling of optical gap solitons.
\newblock {\em Phys. Rev. Lett.}, 81:85--88, Jul 1998.

\bibitem{Hansson2023}
Tobias Hansson, Pedro Parra-Rivas, and Stefan Wabnitz.
\newblock Modeling of dual frequency combs and bistable solitons in third-harmonic generation.
\newblock {\em Communications Physics}, 6(1):59, 2023.

\bibitem{fan2022topological}
Zhiwei Fan, Danila~N Puzyrev, and Dmitry~V Skryabin.
\newblock Topological soliton metacrystals.
\newblock {\em Communications Physics}, 5(1):248, 2022.

\bibitem{lu21}
Juanjuan Lu, Ayed Al~Sayem, Zheng Gong, Joshua~B Surya, Chang-Ling Zou, and Hong~X Tang.
\newblock Ultralow-threshold thin-film lithium niobate optical parametric oscillator.
\newblock {\em Optica}, 8(4):539--544, 2021.


\newblock Hyperparametric solitons in nondegenerate optical parametric oscillators.
\bibitem{ciret2020}
Ciret C, Alexander K, Poulvellarie N, Billet M, Mas Arabi C, Kuyken B, Gorza S P and Leo F 2020 \textit{Optics Express} \textbf{28} 31584--31593
\bibitem{gerini2025}
Gerini A, Ravaro M, Th{\'e}veneau C, Larrue A, Garcia M, Krakowski M, G{\'e}rard B and Leo G 2025 \textit{Optics Express} \textbf{33} 17560--17565

\bibitem{Weng2026hyperparam}
Haizhong Weng, Xinru Ji, Mugahid Ali, Edward~H Krock, Lulin Wang, Vikash Kumar, Weihua Guo, Qing Wan, Tobias~J Kippenberg, John~F Donegan, et~al.






\end{thebibliography}
\end{document}